\documentclass[12pt]{article}
\usepackage[centertags]{amsmath}
\usepackage{amssymb}
\usepackage{amsthm}
\usepackage{enumerate}
\usepackage{url}

\usepackage[english]{babel}
\usepackage[latin1]{inputenc}

\usepackage{color}
\usepackage[normalem]{ulem}

\makeatletter
  \def\tagform@#1{\maketag@@@{(#1)\@@italiccorr}}
\makeatother

\newcommand{\CM}{\mbox{${C^{\infty }(M)}$}}
\newcommand{\CGamma}{\mbox{${C^{\infty }(\Gamma )}$}}
\newcommand{\Cstar}{\mbox{${C^{*}}$}}
\newcommand{\calZ}{\mbox{${\mathcal Z}$}}
\newcommand{\ZA}{\mbox{${\mathcal Z(\calA )}$}}

\newcommand{\DerA}{\mbox{$\mathrm{Der(\calA )}$}}
\newcommand{\calG}{\mbox{${\mathcal G}$}}

\newcommand {\calA}{\mbox{${\mathcal A}$}}

\begin{document}

\title{Creation of Matter in a~Noncommutative Universe}

\author{Tomasz Miller\\
Faculty of Mathematics and Information Science, \\ Warsaw
University of
Technology \\
ul. Koszykowa 75, 00-662 Warsaw, Poland \\
and Copernicus Center for Interdisciplinary Studies,\\ Cracow, Poland
\and Michael Heller \\
Copernicus Center for Interdisciplinary Studies, Cracow, Poland \\
and Vatican Observatory,
V-00120 Vatican City State}

\date{\today}

\maketitle

\begin{abstract}

The dark matter and dark energy problem, that is now dominating the~research in cosmology, makes the~question of the~origin of mass-energy content of the~universe more urgent than ever. There are two philosophies regarding this question: according to Mach's principle it is matter that generates geometry of space-time, and according to Wheeler's geometrodynamics some configurations of space-time geometry are to be interpreted as its material content. Neither of these philosophies has led to success. In the~present paper, we show that there exists an~algebraic generalisation of geometry that reconciles, in a~sense, these two seemingly opposite standpoints. The~geometry is constructed with the~help of a~noncommutative algebra of smooth functions on a~groupoid and its derivations. The~groupoid in question has a~nice physical interpretation: it can be regarded as a~space of Lorentz rotations. In this way, Lorentz symmetries are inherent to the~generalised geometry of space-time. We define the~action for this geometry and, by varying it, obtain generalised vacuum Einstein equations (for a~simplified model). It turns out that these equations contain additional terms (with respect to the~standard vacuum Einstein equations) which are naturally interpreted as the~components of the~energy-momentum tensor. Matter is thus created out of purely geometric degrees of freedom. We find two exact solutions (for even more simplified case). We argue that the~creation of matter, being a~global effect, makes the~contrast between Mach and Wheeler philosophies ineffective.
\end{abstract}
\maketitle

\section{Introduction}
The dark matter and dark energy problem, that is now dominating the~research in cosmology, makes the~question of the~origin of mass-energy content of the~universe more urgent than ever. This problem could hardly be solved without a~deeper understanding of matter creation mechanisms in the~universe. Einstein's equations are plagued with a~dualism of geometry (left hand side of the~equations) and matter (their right hand side). Einstein looked for a~remedy against this difficulty in the~idea called by him Mach's principle. It admits several nonequivalent formulations, some stronger of which claim that (local) geometry should totally be determined by the~global distribution of matter throughout space-time. The~idea could be encapsulated in the~slogan ``geometry out of matter'' \cite{Bucket}. The~opposite philosophy was  propagated by John Archibald Wheeler who argued that the~material sources that appear in Einstein's equations can entirely be reconstructed from a~characteristic imprint they exert on the~space-time geometry. Here we have a~slogan ``matter out of geometry''. The~program was known as ``Wheeler's geometrodynamics''\cite{Wheeler1964,Wheeler1968,Wheeler1973}. The~fact that neither of these philosophies, attractive as they are, has led to the~success suggests that a~stumbling block lies in the~very concept of geometry that is too rigid to accommodate for such a~complex phenomenon as that of matter. In the~present paper, we explore a~generalisation of the~standard differential geometry and, basing on it, construct cosmological models in which two above mentioned philosophies not only seem to work, but also could, in a~sense, be unified.

There are two ways of doing differential geometry: by using local coordinate systems, and algebraically in terms of smooth functions on a~given space, the~latter being independent of the~choice of coordinates. Both these methods are equivalent, but the~second is better suited for generalisations. Let us then consider a~space-time manifold $M$ and the~algebra $C^{\infty}(M)$ of smooth functions on $M$. The~algebra \CM \ can, without losing its geometric properties, be replaced by a~more general, not necessarily commutative, algebra on a~more general space than the~space-time manifold $M$. The~more general space we use to this end is the~so-called groupoid, denoted by $\Gamma $ (for details see Section 2). The~choice of this space has a~nice physical motivation. the~groupoid concept is a~generalisation of the~group concept, and as such it generalizes the~notion of symmetry. In our case, the~elements of the~groupoid $\Gamma $ can be regarded as Lorentz rotations. Therefore, $\Gamma $ can serve as a~more natural environment for general relativity (being the~space of its symmetries) than the~usual ``naked'' space-time.

We then consider the~algebra \CGamma \ of smooth (compactly supported) functions on a~groupoid $\Gamma $. The~consequence of these rather simple replacements is that space-time points acquire their internal structure, i.e. internal degrees of freedom not unlike in the~Kaluza--Klein type models, with the~difference that now internal degrees of freedom are really ``internal'', not created by adding new space dimensions.

The algebra \CGamma \ can easily be made into a~\Cstar-algebra which assimilates the~formalism to that employed in quantum mechanics. In fact, this approach has been used  to construct a~model unifying general relativity and quantum mechanics with a~perspective to make an~attempt at a~quantum gravity theory \cite{HSL97,HS99,HPS05,Heller3}. It goes without saying that truly fundamental mechanism of matter creation can hardly be imagined without the~correctly working quantum gravity theory. This is why the~present work can only be regarded as a~preliminary step in this direction.

Models presented in this paper are nothing more than ``toy models''; however, they seem worthwhile to be explored not only because they give such surprising results as far as matter generation is concerned, but also because they are preparing mathematical tools to deal with more realistic situations. The~present work is based on our research paper \cite{Canadian}.

The plan of our essay runs as follows. In Section 2, we briefly discuss the~method of doing geometry on the~groupoid $\Gamma $ in terms of the~algebra $C^{\infty }(\Gamma )$ and its derivations. In Section 3, we construct such a~geometry for a~groupoid with a~finite structure group. Then, in Section 4, we deduce generalised Einstein equations from the~corresponding action principle, and show that they, when projected onto space-time, contain an~additional term that can naturally be interpreted as a~``matter source''. We also find two explicit ``Friedman-like'' solutions for a~simplified form of the~metric.  Finally, we append some concluding remarks.

\section{Space of Lorentz Symmetries and Its Geometry}
A natural setting for relativity theory is a~set of pairs of reference frames with a~Lorentz transformation acting between them. This setting can be given strict mathematical form. Let $M$ be a~space-time, and let us consider the~bundle of reference frames (frame bundle, for short) on $M$, denoted by $(E, \pi_M)$ where $E$, called the~total space of the~bundle, is the~set of all local reference frames on $M$, and $\pi_M: E \rightarrow M$ is a~mapping projecting a~frame $p \in E$ to its attachment point $x\in M$, $\pi_M(p)=x$. The~set of all frames attached to $x$, $E_x = \pi_m^{-1}(x)$, is called the~fiber at $x$. Let $G$ be a~Lorentz group or one of its subgroups. It acts on $E$ along fibres, i.e. any two frames in the~same fibre can be transformed into each other with the~help of an~element of $G$.\footnote{This means that we consider only Lorentz rotations; translations, i.e. transformations between fibres, are excluded.} In this way, the~Cartesian product has been constructed,
\[
\Gamma = E \times G = \{\gamma = (p,g): p \in E, g \in G\}.
\]
It is clear that Lorentz transformations in a~given fibre can, as elements of a~group, be suitably composed and have inverses. The~construction, just described, is called transformation groupoid with $\Gamma $ as its groupoid space, $E$ its base space, and $G$ its structure group. This purely algebraic construction can be equipped with the~smoothness structure; it is then called smooth transformation groupoid.

Just as geometry of space-time $M$ can be done in terms of the~algebra \CM \ of smooth functions on $M$, a~generalised geometry of space-time $M$ can be done in terms of an~algebra \calA \ on the~groupoid $\Gamma = E \times G$. The~usual algebra $C^{\infty}(\Gamma )$ of smooth (compactly supported) functions on $\Gamma $ (with pointwise multiplication, denoted by $\cdot $) would reproduce geometry of $\Gamma $ with nothing interesting for our program. To obtain an~interesting generalisation we replace the~commutative pointwise multiplication with a~not necessarily commutative multiplication $*$. Its concrete form need not bother us here (for details see \cite{Canadian}). We thus consider the~algebra $ \calA = (C^{\infty}(\Gamma ), * )$ of smooth (compactly supported) functions on $\Gamma $. It is, in general, noncommutative which means that if $a,b \in \calA$ then $a*b \neq b*a$. This seemingly innocuous modification leads to dramatic changes in geometry. We also impose on \calA\ suitable smoothness conditions.

Which is the~contact of the~algebra \calA \ with the~algebra \CM \ of smooth functions on space-time $M$? To answer this question we remind the~concept of the~center of a~noncommutative algebra. The~center of an~algebra $A$, denoted $\calZ (A )$ is the~set of all elements of $A$ that commute with all elements of $A$ (it is obvious that if an~algebra is commutative, it coincides with its center). It can be shown that the~center of our algebra \calA \ is isomorphic with \CM . Therefore, if, in constructing geometry, we restrict the~algebra \calA \ to its center $\calZ(\calA )$, we obtain the~geometry of space-time $M$.

The standard manifold geometry is encoded in its metric tensor. Mathematically, metric is a~function taking two smooth vector fields as its arguments and returning a~smooth function on this manifold. In the~algebraic approach to geometry, the~counterparts of vector fields are derivations of the~corresponding algebra. Let us consider our algebra \calA . Its derivation is a~linear map $v: \calA \rightarrow \calA $ satisfying the~well known Leibniz rule
\[
v(a * b) = v(a)*b + a*v(b)
\]
for $a, b \in \calA $. The~set of all derivations of the~algebra \calA \ is denoted by \DerA .

Taking the~above into account, we assume the~metric of the~form $\calG : V \times V \rightarrow \ZA $ where $V \subseteq \DerA $.\footnote{$V$ should have the~\ZA -module structure as well as the~Lie algebra structure.} The~pair $(\calA , V)$ is called differential algebra. It serves us as the~basic structure to construct an~algebraic version of generalised geometry. We proceed in strict analogy to what is usually done when developing the~standard differential geometry. We construct connection (with the~help of the~Koszul formula), curvature and all other magnitudes necessary to write down Einstein's field equations. In the~following, when pursuing this program, we shall limit ourselves to a~special case of a~finite model.

\section{Geometry on the~Groupoid Algebra with Finite Structure Group}

Let us consider an~$n$-element subgroup $G$ of the~group of Lorentz rotations. Without loss of generality, we can assume that the~frame bundle studied is trivial: $E = M \times G$, and therefore the~considered groupoid is
\begin{align*}
\Gamma = M \times G \times G = \{ (x,g_i,g_j) : x \in M, \ g_i,g_j \in G, \ i,j = 1,\ldots,n \}.
\end{align*}
The~noncommutative multiplication $\ast$ in the~algebra $C^{\infty}(\Gamma)$ of smooth (compactly supported) functions on $\Gamma$ is given by
\begin{align*}
\forall \, a,b \in C^{\infty}(\Gamma) \quad (a \ast b)(x, g_k ,g_l) := \sum\limits_{m=1}^{n} a(x, g_k, g_m) b(x, g_k g_m, g_m^{-1} g_l).
\end{align*}
It is much more convenient, however, to regard every function $a \in C^{\infty}(\Gamma)$ as an~$n$-by-$n$ \emph{matrix} $(a_{ij})$ with $C^\infty(M)$-valued entries, defined via
\begin{align*}
a_{ij}(x) := a(x, g_i, g_i^{-1} g_j).
\end{align*}
In this way, the~algebra studied becomes
\begin{align*}
{\cal A}_n := \mathbb{M}_n(C^\infty(M)) = C^{\infty}(M) \otimes \mathbb{M}_n(\mathbb{C})
\end{align*}
\noindent
and the~operation $\ast$ becomes nothing but the~standard matrix multiplication
\begin{align*}
\forall \, a,b \in {\cal A}_n \quad (a \ast b)_{ij}(x) := \sum\limits_{k=1}^{n} a_{ik}(x) b_{kj}(x).
\end{align*}
Let us note that algebras similar to ${\cal A}_n$ commonly appear in the~exploration of possible applications of noncommutative geometry to physics, most notably in the~context of the~Noncommutative Standard Model of particle physics (see \cite{Connes,Chamseddine} for details on noncommutative geometry and the~Noncommutative Standard Model or \cite{Connes2,NCSM} for a~more accessible review).

The center ${\cal Z}({\cal A}_n)$ of this algebra consists of matrices of the~form $f \, I_n$, where $f \in C^\infty(M)$ and $I_n$ denotes the~$n$-by-$n$ identity matrix. It is, therefore, isomorphic to the~algebra $C^\infty(M)$, just as the~previous section anticipated. From now on we shall identify ${\cal Z}({\cal A}_n)$ with $C^\infty(M)$.

We consider the~full $C^\infty(M)$-module of derivations of the~algebra ${\cal A}_n$, $V := \textnormal{Der} \, {\cal A}_n$. It can be shown that $V$ decomposes into the~direct sum of two its submodules:
\begin{itemize}
\item The~submodule $\textnormal{Hor} \, {\cal A}_n$ of \emph{horizontal} derivations. Its elements are \emph{liftings} of the~smooth vector fields on the~manifold $M$ onto ${\cal A}_n$. More explicitly, for any smooth vector field $X \in \textnormal{Der} \, C^\infty(M)$ one defines its lifting as a~map $\bar{X}: {\cal A}_n \rightarrow {\cal A}_n$ acting entrywise, namely
\begin{align*}
\forall \, a \in {\cal A}_n \quad (\bar{X} a)_{ij} := X a_{ij}.
\end{align*}
\item The~submodule $\textnormal{Inn} \, {\cal A}_n$ of \emph{inner} derivations. By an~\emph{inner derivation} induced by an~element $b \in {\cal A}_n$ one understands a~map $\textnormal{ad}_b: {\cal A}_n \rightarrow {\cal A}_n$ defined as
\begin{align*}
\forall \, a \in {\cal A}_n \quad \textnormal{ad}_b a := [b,a] = b \ast a - a \ast b.
\end{align*}
\end{itemize}

Embarking on the~construction of the~generalised geometry on the~differential algebra $({\cal A}_n, V)$, one begins with the~\emph{metric} $\calG : V \times V \rightarrow \ZA $, upon which the~Levi-Civita connection and the~curvature tensors are subsequently defined. All these objects can be, and usually are, studied by physicists with the~help of the~so-called \emph{abstract-index notation}. The~idea behind this notation is to identify the~abstract (usually tensorial) objects with the~array of their components in a~chosen basis of $V$. Multi-indexed expressions, obtained in this way, can then be effectively manipulated under the~set of simple rules, among others including the~Einstein summation convention  (see \cite[Chapter 2]{PenroseRindler}).

The natural choice for the~(local) basis of $\textnormal{Hor} \, {\cal A}_n$ is the~lifting of the~\emph{coordinate basis} $\left(\overline{\tfrac{\partial}{\partial x^\mu}}\right)$ induced by some chart $x$. In the~following, we shall denote the~liftings of the~coordinate vector fields $\overline{\tfrac{\partial}{\partial x^\mu}}$ simply by $\partial_\mu$, suppressing both the~overline and the~reference to the~inducing chart and using also other \emph{lowercase Greek letters} $\nu, \lambda, \sigma, \ldots$. We adopt the~usual convention that these indices can assume values $0,1,\ldots,m-1$, where $m = \dim M$.

On the~other hand, one can show that any basis of $\textnormal{Inn} \, {\cal A}_n$ contains exactly $n^2-1$ elements. In what follows, we shall not specify the~basis concretely, but we shall denote its elements by $\partial_{\{A\}}$ (by analogy with the~horizontal derivations), using also other \emph{capital Latin letters in curly brackets} $\{B\},\{C\},\{D\},\ldots$. These indices can assume values $\{1\},\{2\},\ldots,\{n^2-1\}$ (interpreted as the~``inner degrees of freedom''). The~use of curly brackets assures that we do not mix the~values of indices of the~two types.

Summarizing, the~basis of $V$ contains two kinds of elements: the~horizontal derivations and the~inner derivations, and this is reflected in the~two types of indices used: the~lowercase Greek indices and the~capital Latin indices in curly brackets. Additionally, it will be convenient to write $\partial_A$ for a~generic derivation from the~basis, using also other \emph{capital Latin letters} $B,C,D,\ldots$. These indices can take all values assumed by the~indices of the~two types listed above.

Let us note here that the~Einstein summation convention applies to each of the~three types of indices separately.
\\

The noncommutativity of ${\cal A}_n$ has a~direct effect on the~commutation relations between $\partial_A$'s. Even though $[\partial_\mu, \partial_\nu] = 0$, exactly as for the~coordinate basis in the~standard differential geometry, in general we have that $[\partial_A, \partial_B] \neq 0$. Mathematically speaking, some of the~\emph{structure constants} $\textbf{c}_{AB}^{\ \ \ \, C}$, defined by the~formula $[\partial_A, \partial_B] = \textbf{c}_{AB}^{\ \ \ \, C} \partial_C$, are nonzero.

Having specified the~basis, we can now construct the~generalised geometry on the~differential algebra $({\cal A}_n, V)$. In the~abstract-index notation, the~metric $\calG$ is represented by the~doubly indexed array $g_{AB}$ where
\begin{align*}
g_{AB} := {\cal G} ( \partial_A, \partial_B ).
\end{align*}
One can regard $g_{AB}$ as a~square, symmetric and nonsingular matrix of order $m + n^2 - 1$. Its inverse matrix is denoted by $g^{AB}$. Exactly as in the~standard differential geometry, the~metric matrix and its inverse can be used to lower and raise indices of other multi-indexed entities.

We are now ready to define the~\emph{Levi-Civita connection} $\nabla: V \times V \rightarrow V$ by means of the~Koszul formula. Namely, for any $u,v \in V$, $\nabla_u v$ is the~unique derivation which satisfies
\begin{align*}
\calG \left(\nabla_u v, w\right) & := \tfrac{1}{2} \big[ u\left({\cal G} (v,w)\right) + v\left({\cal G} (u,w)\right) - w\left({\cal G} (u,v)\right) \big.
\\
& \ + \big. {\cal G}(w,[u,v]) + {\cal G}(v,[w,u]) - {\cal G}(u, [v,w]) \big].
\end{align*}
\noindent
for all $w \in V$. The~connection's components (called the~Christoffel symbols of the~second kind) are defined by the~equality
\begin{align*}
\nabla_{\partial_C} \partial_B = \Gamma^A_{\ BC} \partial_A
\end{align*}
\noindent
and by the~Koszul formula they can be expressed as
\begin{align*}
\Gamma^A_{\ BC} = \tfrac{1}{2}g^{AD} \left( \partial_C g_{DB} + \partial_B g_{DC} - \partial_D g_{BC} + \textbf{c}_{CBD} + \textbf{c}_{DCB} - \textbf{c}_{BDC} \right).
\end{align*}
Due to the~noncommutativity, we have
\begin{align*}
\Gamma^A_{\ BC} - \Gamma^A_{\ CB} = \textbf{c}^{\ \ \ \, A}_{CB}.
\end{align*}
Therefore, unlike the~components of the~standard Levi-Civita connection, $\Gamma^A_{\ BC}$ might not be symmetric with respect to the~latter two indices.

In spite of this asymmetry, the~Levi-Civita connection enjoys many of the~properties of its standard counterpart, in particular, it is \emph{torsion-free}, that is
\begin{align*}
\forall \, u,v \in V \quad \nabla_u v - \nabla_v u - [u, v] = 0
\end{align*}
\noindent
and \emph{compatible with the~metric}, that is
\begin{align*}
\forall \, u,v,w \in V \quad w\left({\cal G} (u,v)\right) = {\cal G} \left( \nabla_w u , v \right) + {\cal G} \left( u, \nabla_w v \right)
\end{align*}
Moreover, just like in the~standard case, the~Levi-Civita connection is the~unique map $V \times V \rightarrow V$ satisfying the~above two conditions.

The~\emph{Riemann curvature tensor} $R: V \times V \times V \rightarrow V$, $(u,v,w) \mapsto R(u,v)w$ is defined by means of $\nabla$ via
\begin{align*}
R(u,v)w := \nabla_u \nabla_v w - \nabla_v \nabla_u w - \nabla_{[u,v]}w.
\end{align*}
Its components $R^C_{\ DAB}$ are defined through
\begin{align*}
R(\partial_A, \partial_B) \partial_D = R^C_{\ DAB} \partial_C,
\end{align*}
\noindent
and are given by the~formula
\begin{align*}
R^C_{\ DAB} = \partial_A \Gamma^C_{\ DB} - \partial_B \Gamma^C_{\ DA} + \Gamma^K_{\ DB} \Gamma^C_{\ KA} - \Gamma^K_{\ DA} \Gamma^C_{\ KB} - \textbf{c}_{AB}^{\ \ \ \, K} \Gamma^C_{\ DK}.
\end{align*}

This is nothing but the~standard expression for the~components of the~Riemann tensor with an~additional term that can be regarded as due to noncommutativity. It can be easily checked that $R$ enjoys the~usual Riemann tensor symmetries
\begin{align*}
& R_{CDAB} = - R_{DCAB} = - R_{CDBA} = R_{ABCD},
\\
& R^C_{\ DAB} + R^C_{\ BDA} + R^C_{\ ABD} = 0.
\end{align*}

Finally, the~Ricci tensor, $\textbf{ric}: V \times V \rightarrow C^\infty(M)$ and the~curvature scalar $r \in C^\infty(M)$ can be introduced as suitable contractions of the~Riemann tensor. Concretely, the~components of the~Ricci tensor read
\begin{align*}
\textbf{ric}_{AB} := R^C_{\ ACB},
\end{align*}
\noindent
whereas the~curvature scalar is
\begin{align*}
r := g^{AB} \textbf{ric}_{AB} = g^{AB} R^C_{\ ACB}.
\end{align*}
Notice that, just as in the~standard case, $\textbf{ric}_{AB} = \textbf{ric}_{BA}$.

With the~generalised curvature tensors defined, we are ready to formulate and study the~generalised vacuum Eistein equations.

\section{Generalised Einstein Equations}

The standard derivation of the~Einstein equations is conducted by means of the~action principle, starting from a~suitably chosen action functional. In standard GR, the~so-called Einstein--Hilbert action is the~integral of the~curvature scalar over the~entire space-time manifold. Therefore, the~natural candidate for the~generalised Einstein--Hilbert action is
\begin{align}
\label{action}
S_{EH} := \int r \sqrt{|g|} \, d^m x,
\end{align}
\noindent
where $m = \dim M$ and $g$ denotes the~determinant of the~metric matrix $g_{AB}$. We want to study the~vacuum equations and therefore postulate no additional matter term.

The action principle amounts here to varying $S_{EH}$ with respect to $\delta g^{A B}$ and assuming that the~variation vanishes. The~calculation leads to the~generalised Einstein equations of the~form
\begin{align}
\label{Einstein_eqs0}
\textbf{ric}_{A B} = 0.
\end{align}

Although these equations look similar to the~standard vacuum Einstein equations, they actually have a~far richer content when projected onto space-time $M$. This is due to the~extra terms coming from additional components of the~metric (the ``inner degrees of freedom''). These extra terms can be interpreted as an~``$m$-dimensional matter--energy'' induced by the~generalised (vacuum) Einstein equations (similarly as it is the~case in the~Kaluza--Klein-type theories \cite{Book1,Book2,Wesson,Clifton}, although without introducing extra geometrical dimensions). Let us consider an~example of the~following block diagonal metric
\begin{align*}
g_{A B} =
\left[\begin{array}{cc}
    g_{\mu \nu} & 0 \\
    0 & g_{\{A\} \{B\}}
  \end{array}\right].
\end{align*}
\noindent
Let us remember that $g_{\mu \nu}$ is an~$m$-by-$m$ matrix and $g_{\{A\} \{B\}}$ is a~$\left(n^2-1\right)$-by-$\left(n^2-1\right)$ matrix, where $n = |G|$. The~inverse metric matrix $g^{A B}$ is also block diagonal.

Such block diagonal metric has an~important feature: its corresponding Levi-Civita connection $\nabla$ \emph{extends the~standard Levi-Civita connection} $\widetilde{\nabla}: \textnormal{Der} \, C^\infty(M) \times \textnormal{Der} \, C^\infty(M) \rightarrow \textnormal{Der} \, C^\infty(M)$ in the~sense that
\begin{align*}
\forall \, \bar{X}, \bar{Y} \in \textnormal{Hor} \, {\cal A}_n \quad \nabla_{\bar{X}} \bar{Y} = \overline{\widetilde{\nabla}_X Y}.
\end{align*}
In other words, the~Levi-Civita connection acts on horizontal derivations in the~same way as does its classical counterpart.

For the~block diagonal metric, the~generalised Einstein equations can be written more explicitly than (\ref{Einstein_eqs0}) as
\begin{align}
\label{Einstein_eqs1}
& \widetilde{\textbf{ric}}_{\mu \nu} = \tfrac{1}{4} \left( \widetilde{\nabla}_\mu \widetilde{\nabla}_\nu \ln |\breve{g}| + g^{\{A\} \{B\}} \widetilde{\nabla}_\mu \widetilde{\nabla}_\nu g_{\{A\} \{B\}} \right),
\\
\label{Einstein_eqs2}
& \textbf{c}_{\{B\}}^{ \quad \{C\}\{D\}} \partial_\mu g_{\{C\}\{D\}} = 0,
\\
\begin{split}
\label{Einstein_eqs3}
& \widetilde{\Delta} g_{\{A\} \{B\}} - g_{\{A\} \{C\}} g_{\{B\} \{D\}} \widetilde{\Delta} g^{\{C\} \{D\}} + \partial^\mu \ln | \breve{g} | \partial_\mu g_{\{A\} \{B\}}
\\
& \quad = - 2 \textbf{c}_{\{A\}}^{ \quad \{C\}\{D\}} \left( \textbf{c}_{\{B\}\{C\}\{D\}} + \textbf{c}_{\{B\}\{D\}\{C\}} \right) + \textbf{c}^{\{C\}\{D\}}_{\qquad \ \{A\}} \textbf{c}_{\{C\}\{D\}\{B\}}.
\end{split}
\end{align}
\noindent
where
\begin{itemize}
  \item $\widetilde{\textbf{ric}}_{\mu \nu}$ denotes the~standard Ricci tensor.
  \item $\widetilde{\nabla}_\mu$ is the~covariant derivative resulting from the~standard Levi-Civita connection $\widetilde{\nabla}$. Note that $\widetilde{\nabla}_\mu$ by definition ``sees'' only the~lowercase Greek indices.
  \item $\widetilde{\Delta} := g^{\mu \nu} \widetilde{\nabla}_\mu \widetilde{\nabla}_\nu$ is the~standard Laplace--Beltrami operator.
  \item $\breve{g} := \det g_{\{A\} \{B\}}$.
\end{itemize}

Equation (\ref{Einstein_eqs1}) is a~projection of the~generalised Einstein equations (\ref{Einstein_eqs0}) onto the~$m$-dimensional space-time $M$. Since it implies that
\begin{align}
\label{Einstein_eqs4}
& \widetilde{r} = \tfrac{1}{4} \left( \widetilde{\Delta} \ln |\breve{g}| + g^{\{A\} \{B\}} \widetilde{\Delta} g_{\{A\} \{B\}} \right),
\end{align}
\noindent
where $\widetilde{r}$ denotes the~standard curvature scalar, therefore we can equivalently write (\ref{Einstein_eqs1}) in the~form of the~standard Einstein equations with a~certain nonzero energy--momentum tensor
\begin{align}
\begin{split}
\label{Einstein_eqs5}
\widetilde{G}_{\mu \nu} & = \tfrac{1}{4} \left[ \left( \widetilde{\nabla}_\mu \widetilde{\nabla}_\nu - \tfrac{1}{2} g_{\mu \nu} \widetilde{\Delta} \right) \ln |\breve{g}| \right.
\\
& \left. \quad \quad + \, g^{\{A\} \{B\}} \left( \widetilde{\nabla}_\mu \widetilde{\nabla}_\nu - \tfrac{1}{2} g_{\mu \nu} \widetilde{\Delta} \right) g_{\{A\} \{B\}} \right],
\end{split}
\end{align}
\noindent
where $\widetilde{G}_{\mu \nu} := \widetilde{\textbf{ric}}_{\mu \nu} - \tfrac{1}{2} g_{\mu \nu} \widetilde{r}$ is the~standard Einstein tensor.

The fact that there appears a~nonzero energy--momentum tensor can be regarded as a~realisation of the~``matter out of geometry'' mechanism \cite{Heller3} or, in this case more precisely, of the~``scalar fields out of noncommutative geometry'' mechanism. One can thus regard equations (\ref{Einstein_eqs2}, \ref{Einstein_eqs3}) as the~equations of state of those ``emergent'' scalar fields.

Additionally, we can also rewrite the~generalised Einstein--Hilbert action (\ref{action}) more explicitly. Namely
\begin{align*}
S_{EH} = \int \left[ \sqrt{|\breve{g}|} \widetilde{r} - \tfrac{1}{4} \sqrt{|\breve{g}|} \left( \widetilde{\Delta} \ln |\breve{g}| + g^{\{A\} \{B\}} \widetilde{\Delta} g_{\{A\} \{B\}} + \textbf{C} \right) \right] \sqrt{-\tilde{g}} \, d^m x,
\end{align*}
\noindent
where
\begin{itemize}
  \item $\tilde{g} := \det g_{\mu \nu}$.
  \item $\textbf{C} := \textbf{c}^{\{B\}\{C\}\{D\}} \left( 2 \textbf{c}_{\{B\}\{D\}\{C\}} + \textbf{c}_{\{B\}\{C\}\{D\}} \right)$.
\end{itemize}
Notice that noncommutativity enters into the~action both through the~additional compoments of the~metric (i.e. the~``inner derivation block'' $g_{\{A\} \{B\}}$) as well as through the~nonzero structure constants.

One can regard the~theory, obtained in this way, as an~example of a~scalar-tensor theory\footnote{See e.g. \cite{Clifton} and references therein.}, which involves no less than $n^2(n^2-1)/2$ independent scalar fields. These fields, when arranged into a~symmetrical matrix $g_{\{A\}\{B\}}$, are such that $\breve{g} := \det g_{\{A\}\{B\}}$ is a~nowhere vanishing field. Notice, moreover, that $\sqrt{|\breve{g}|}$ seems to play a~special role in this theory, as it can be shown to satisfy the~following Klein--Gordon-like equation
\begin{align}
\label{Einstein_eqs6}
& \left( \widetilde{\Delta} + \tfrac{1}{4} \textbf{C} \right) \sqrt{|\breve{g}|} = 0
\end{align}
\noindent
with $\tfrac{1}{4}\textbf{C}$ playing the~role of the~mass term. This might constitute another illustration of how noncommutative geometry gives rise to (massive) scalar fields in this model.
\\

We now move to presenting two explicit solutions of Einstein equations (\ref{Einstein_eqs1}-\ref{Einstein_eqs3}) in the~simplest case when the~structure group $G$ has only two elements. We restrict ourselves to ``Friedman-like'' solutions, by which we mean metrics, whose ``horizontal derivation block'' $g_{\mu \nu}$ is of the~Friedman--Lema\^{\i}tre--Robertson--Walker form and whose remaining components are time-dependent only\\footnote{As usual, we set $c=G=1$.}
\begin{align*}
& g_{\mu \nu} =
\left[\begin{array}{ccccccc}
    -1 & 0 & 0 & 0 \\
    0 & \frac{a^2(t)}{1 - k r^2} & 0 & 0\\
    0 & 0 & a^2(t) r^2 & 0 \\
    0 & 0 & 0 & a^2(t) r^2 \sin^2 \theta
\end{array}\right],
\\
& g_{\{A\} \{B\}} = g_{\{A\} \{B\}}(t).
\end{align*}
\noindent
Recall that $a(t)$ is called the~\emph{scale factor} and $k \in \{-1,0,1\}$ is the~\emph{curvature constant}.

To simplify the~calculations, we assume furthermore that the~``inner derivation block'' $g_{\{A\} \{B\}}$ has the~following form
\begin{align*}
g_{\{A\} \{B\}}(t) =
\left[\begin{array}{ccc}
    \xi f^2(t) & 0 & 0\\
    0 & 0 & \eta f^2(t)\\
    0 & \eta f^2(t) & 0
  \end{array}\right],
\end{align*}
\noindent
where $\xi, \eta \in \mathbb{R} \setminus \{0\}$ and $f$ is a~time-dependent nowhere vanishing function.

One of the~Einstein equations, (\ref{Einstein_eqs2}), is satisfied automatically. The~remaining two equations (\ref{Einstein_eqs1}, \ref{Einstein_eqs3}) yield the~following overdetermined nonlinear system of ordinary differential equations
\begin{align}
\label{ODE}
\left\{ \begin{array}{rl}
    \frac{\ddot{a}}{a} + \frac{\ddot{f}}{f} & = 0,
    \\
    a \ddot{a} f + 2 \dot{a}^2 f + 3 a \dot{a} \dot{f} & = - 2 k f,
    \\
    f \ddot{f} a + 2 \dot{f}^2 a + 3 f \dot{f} \dot{a} & = \frac{1}{\eta} a.
  \end{array}\right.
\end{align}
\noindent
together with an~additional algebraical condition that $\xi = 2 \eta$.

Notice that, by the~last of the~above equations the~function $f$ cannot be constant. Moreover, the~Hubble parameter $H$ can be expressed entirely in terms of $f$ and $\eta$ via
\begin{align*}
H := \frac{\dot{a}}{a} = \frac{\eta^{-1} - f \ddot{f} - 2 \dot{f}^2}{3 f \dot{f}}.
\end{align*}

Two explicit solutions of system (\ref{ODE})\footnote{There might be more, however finding them (or proving that other solutions do not exist) seems like a~daunting task!}, one for $k=0$ and another for $k=-1$, read:
\begin{align}
\label{FLRW_sol1}
& \textnormal{for } k=0 && a(t) = a_0, && f(t) = \tfrac{1}{\sqrt{2\eta}} (t - t_0),
\\
\label{FLRW_sol2}
& \textnormal{for } k=-1 && a(t) = \sqrt{\tfrac{2}{5}} (t - t_0), && f(t) = \tfrac{1}{\sqrt{5\eta}} (t - t_0),
\end{align}
\noindent
where $\eta, a_0, t_0$ are constants. In fact, without any loss of generality one can take $\eta=1$ and $t_0 = 0$. Note that $g_{AB}$ becomes degenerate at $t=0$.

First solution (\ref{FLRW_sol1}) describes the~flat Minkowski space--time, although it is \emph{not} a~static solution since $\dot{f} \neq 0$ (and the~above mentioned degeneracy occurs at $t=0$).

Second solution (\ref{FLRW_sol2}) describes a~hyperbolic, linearly expanding universe with the~initial singularity at $t=0$, resembling (but different from) the~metric studied first by Milne \cite{Milne}, in which $k=-1$ but $a(t) = t$. Milne's unconventional cosmology has recently gained a~renewed interest in the~work of Benoit-L\'{e}vy and Chardin \cite{DiracMilne} in the~form of the~so-called Dirac--Milne universe, which is argued to be a~viable alternative to the~$\Lambda$CMD model. What is noteworthy, both the~Milne model and the~model governed by (\ref{FLRW_sol2}) are free from the~cosmic age problem and from the~horizon problem, where the~latter is solved without introducing inflation \cite{DiracMilne}.

However, there is a~significant difference between Milne's metric and solution (\ref{FLRW_sol2}). Namely, Milne's universe is devoid of energy--matter; it is a~vacuum solution. On the~other hand, solution (\ref{FLRW_sol2}) is associated to a~nonzero energy--matter tensor. Since the~solution is Friedman-like, the~energy--matter tensor describes some sort of a~perfect fluid. Let us therefore see, \emph{what kind of perfect fluid is in this case induced by the~(noncommutative) geometry}.

The~standard Einstein tensor assumes here the~following (mixed) form
\begin{align*}
& \widetilde{G}^\mu_{\ \nu} = \tfrac{3}{2 t^2} \, \textnormal{diag}(3,1,1,1)
\end{align*}

Therefore, by Einstein equations, the~induced perfect fluid energy density $\rho$ and pressure $p$ are
\begin{align*}
\rho(t) = - \frac{36 \pi}{t^2}, \qquad p(t) = \frac{12 \pi}{t^2}.
\end{align*}

Note that $\rho$ is negative and therefore in this case no classical fluid matches the~induced one. In fact, such fluid would violate various energy conditions of general relativity.

Nevertheless, Friedman cosmology involving negative $\rho$ has been studied in \cite{Nemiroff2005}, where it is argued that some realisations and extensions of quantum field theories do allow for such exotic energy forms, albeit only locally. In the~article cited, the~perfect fluid with $\rho < 0$ and $w = p/\rho = -1/3$ is referred to as \emph{negative cosmic strings}. The~interplay between such \emph{negative} forms of energy and the~classical \emph{positive} ones leads to interesting cosmological scenarios, also studied in \cite{Nemiroff2005}.

\section{Concluding remarks}
How do the~models constructed in the~present work inscribe into philosophies (alluded to in the~Introduction) concerning the~relationship between matter and space-time geometry? It is straightforward that they nicely fall into the~``matter out of geometry'' heading.  Original Wheeler's program could not succeed since the~standard space-time geometry had not enough degrees of freedom to accommodate his postulates. As our work has demonstrated, noncommutative generalisation of geometry creates such possibilities. The~remarkable fact is that not only the~energy-momentum tensor is recovered from purely geometric degrees of freedom (without postulating additional space dimensions), but also suitable equations of state can be obtained in this way. The~fact that these astonishing properties are produced in the~framework of very simplified models allows us to expect even more interesting effects when more realistic models are constructed.

The main feature of noncommutative generalisation of geometry consists of a~strong shift in the~interplay between local and global geometric properties. In strongly noncommutative spaces, local properties are entirely engulfed by their global structure. Such spaces are nonlocal entities, in the~sense that local notions in them are, in general, devoid of meaning and can only be recovered as some ``limiting cases''. Our algebra \calA \ of smooth compactly supported functions on the~groupoid $\Gamma $ of Lorentz symmetries with noncommutative multiplication determines a~noncommutative space with milder properties. Local properties are virtually present in it as encoded in the~fact that the~center \ZA \ of the~algebra \calA \ is nontrivial. Owing to this fact, the~usual geometry of space-time, with all its local properties, can naturally be recovered.

And here we have a~contact with Mach's principle. As we remember, according to its strong version all local properties of space-time should be entirely determined by space-time global structure. This is exactly what happens in our models. On the~level of the~algebra \calA , the~global structure is dominating the~scene, and it is this structure that determines, through the~center of \calA , all local properties of space-time $M$. We could say that the~noncommutative space  as determined by the~algebra \calA \ is fully Machian, and all anti-Machian properties of space-time $M$ emerge when ``noncommutative symmetries'' are broken to the~usual space-time symmetries.

In this way, ``ugly dichotomy'' of space-time and matter is removed, and Einstein's starving for a~monistic vision of the~universe could be satisfied, albeit on the~level of simplified models.

\end{document}